\documentclass{ws-ijmpd}

\begin{document}

\markboth{Sigismondi}
{Picard Satellite for Solar Astrometry}

%
\catchline{}{}{}{}{}
%

\title{THE PICARD SATELLITE MISSION FOR SOLAR ASTROMETRY}

\author{Costantino Sigismondi}

\address{ICRA and Sapienza University of Rome (IT), Nice University (FR) and IRSOL Locarno Solar Research Institute (CH)\\
sigismondi@icra.it}

\maketitle

\begin{history}
\received{1 March 2011}
\revised{Day Month Year}
\comby{Managing Editor}
\end{history}

\begin{abstract}

The Picard solar satellite has been launched on June 15, 2010. This mission is dedicated to the measurement of the solar diameter with an expected accuracy of milliarcseconds of arc. The radiometer onboard is to measure the total solar irradiance. The final goal is the evaluation of the W, the logarithmic ratio of radius and luminosity. This parameter will help the climatologists to recover past values of the solar luminosity when the radius is available from ancient eclipses data.

\end{abstract}


\section{Introduction: Current Space Solar Missions}	
There are several space missions dedicated to the Sun currently in flight\cite{Vial2010}. I will consider only the missions with some astrometric interest. SDO is the Solar Dynamics Observatory\footnote{http://sdo.gsfc.nasa.gov}, and it is dedicated to understand the magnetic field in the Sun and the mechanism which drives its 11-years quasi periodic cycle (22 including spots polarity). The Helioseismic and Magnetic Imager (HMI) is of particular interest for solar astrometry. This experiment is onboard of SDO and it is a follow-up of the SOHO\footnote{http://sohowww.nascom.nasa.gov/} MDI experiment, with a better space resolution. The full solar disk covers a 4096x4096 CCD, with a space resolution of 0.5 arcsec per pixel.
The SOHO MDI experiment reported a solar radius which is not changing within 0.01\% \cite{Kuhn2010} during a whole solar cycle, but this accuracy is probably too much optimistic with respect to the errors induced by the thermal adjustements of the instrument, not expressely designed for high precision astrometry.

The PROBA satellites\footnote{http://www.esa.int/SPECIALS/Proba/index.html} are part of ESA's In-orbit Technology Demonstration Programme: missions dedicated to flying innovative technologies. PROBA stands for PRoject for OnBoard Autonomy. The PROBA satellites are among the smallest ever to be flown by ESA, but they are making a big impact in space technology.  PROBA 2 has also observed the partial eclipse of Jan 4, 2011 with its extreme-ultraviolet telescope (SWAP) using new pixel sensor technology (APS), to make measurements of the solar corona in a very narrow band, with the Centre Spatial de Li\`e ge as lead institute supported by the Royal Observatory of Belgium and with a robust industrial team. The two satellites of PROBA 3 will fly in formation (expected in 2015-2016) to form a coronagraph in space, one satellite eclipsing the Sun to allow the second at 150 m of distance to study the otherwise invisible solar corona. S. Koutchmy and C. Bazin at the Institute of Astrophysique de Paris are involved in this study of long-baseline coronograph. The HERSCHEL (Helium Resonant Scattering in the Corona and Heliosphere) investigation successfully obtained unprecedented images of the helium and hydrogen components of the solar corona out to 3 solar radii during a suborbital flight on 14 September 2009 \cite{Herschel}.

RHESSI\footnote{http://hesperia.gsfc.nasa.gov/hessi/index.html}, the Reuven Ramaty High-Energy Solar Spectroscopic Imager satellite is dedicated to the study of the solar flares, since 2002. Its pointing system has been serendipitously exploited to study the solar oblateness.
The solar oblateness changes along the solar cycle according to the measurements of SDS (see after) SOHO MDI and REHSSI, in phase with the solar cycle activity.\cite{Fivian2008}

\section{SDS a Balloon-borne Solar Mission}	

There are also two solar balloon borne missions which observed the Sun from above the 99 \% of the turbulent atmosphere (Solar Disk Sextant\cite{Egidi2006} and SunRise\cite{Sunrise2011}). The projects based on NASA long-duration balloons like SunRise or NASA-Yale University like SDS have a cost much less than space missions, especially now that the ESRANGE base at 67 degrees of latitude North in Sweden is available, where around the summer solstice the Sun is visible without interruption. The use of Anctartica bases implies larger costs.
New Mexico facilities of Fort Sumner have been used for SDS flights (1992, 1994, 1995, 1996 and 2009) as well as for SunRise gondola test flight.
The main difference between space satellites and stratosphere balloons is the gravity, which acts on the latter.
Gravity can modify the curvature of the optics depending on the solar altitude.
Zonal winds carry the balloon towards its final destination, at about 30 km per hour, westward for Northern emisphere.
During the rising phase, about 3 hours, the temperature of the instruments, when no windshield is provided, drops to -40 degrees Celsius, crossing the tropopause layer. The dark parts of the payload (solar arrays) attain even 100 degrees Celsius on the top of stratosphere at 37 km. Thermal gradients on the beam spitting wedge can contribute to the errors, but it is not demostrated that they are determinant\cite{Righini2004}.
The Solar Disk Sextant flights cover almost two cycles of solar activity. The last flight (october 2009) is still under data analysis. The philosophy of this is experiment is strictly related with the Picard satellite mission, in order to extend its scientific goals back and forth well beyond the lifetime of this satellite (for Picard 3 years of activity are expected: 2010-2013).
SDS is devoted exclusively to the measurement of the solar diameter at a 0.01 arcsecond level, while the measurements of the total solar irradiance are already available from other satellite missions\cite{Krivova2011}, since 1978.
The flights can be repeated during the mission of Picard and after, in order to recover accurately the value of the solar diameter and to calibrate the measurement with the satellite data.

\section{Solar Variability: Total Solar Irradiance and Radius Variations}	

The Sun is a star which is experiencing the main sequence phase of its evolution, therefore it is stable over a very long time span.
A reference evolutionary model for a solar mass star calculated by L. Siess (1999) on his website of Li\`ege University\footnote{http://www-astro.ulb.ac.be/~siess/MODELS/PMS/OV02/m1.0z02d02.hrd} and used to prepare the figure 1. 
The standard solar model is used as a test case for the stellar evolution calculation because the luminosity, radius, age and composition of the Sun are well determined (Guenter, 2010)\footnote{http://www.ap.stmarys.ca/\~\ guenther/evolution/what\_is\_ssm.html}. Nevertheless certain properties of the Sun are observed to vary during the course of a sunspot cycle.
An enormous amount of energy would be required to change the whole structure of the Sun,\cite{Callebaut2002} namely the height of the convective envelop, but rather small variations of the surface layer are allowed from the available energy balance.

\begin{figure}[top]
\centerline{\psfig{file=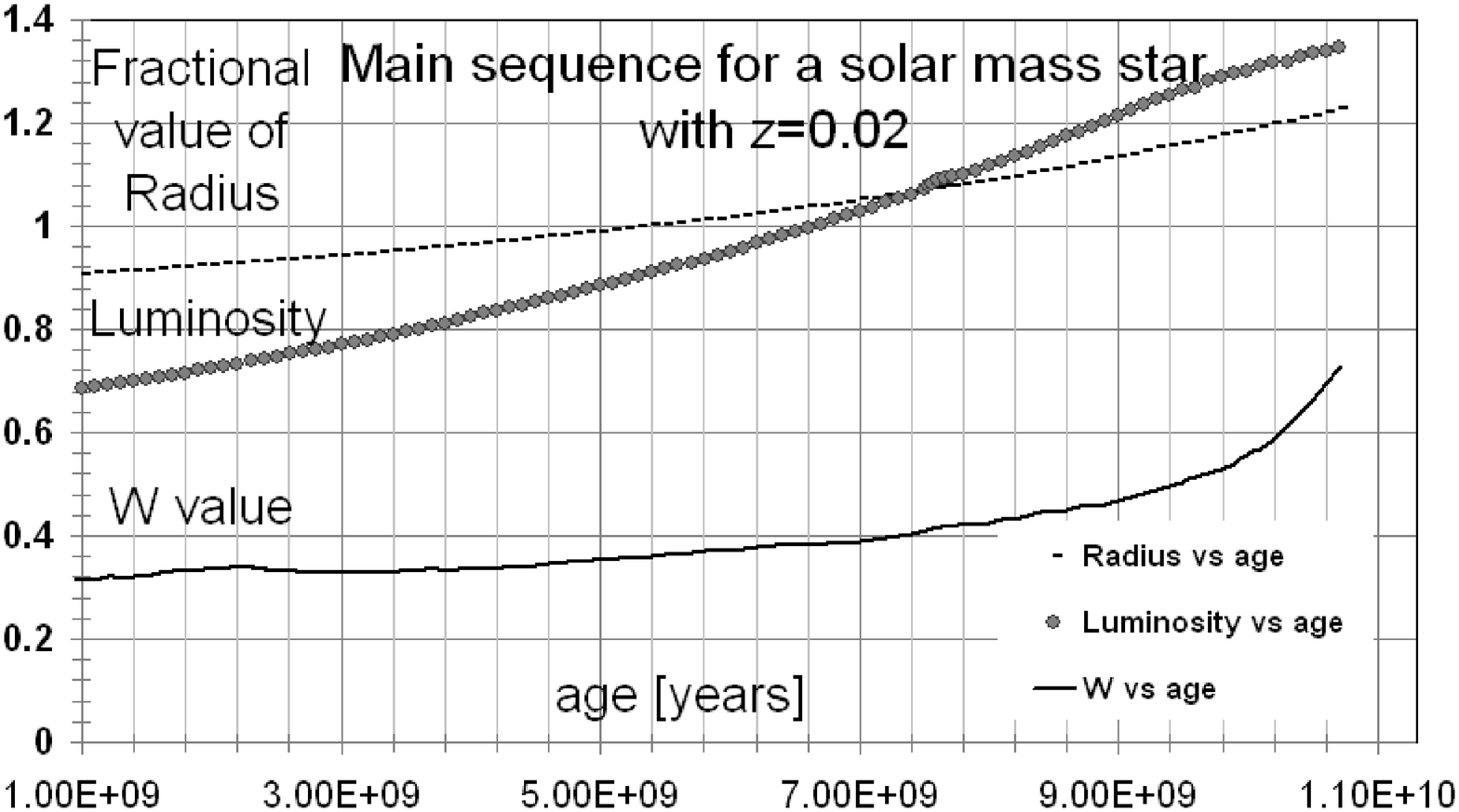,width=13cm}}
\vspace*{0pt}
\caption{This plot represents the evolution of a solar mass star with z=0.02 and overshooting parameter d=2 for pressure scale heights (see L. Seiss (1999)). The Sun has Z/X = 0.02497, and this explains why radius and luminosity do not cross each other at 1 value for present age (4.52 billion years). I have added the plot of W=dlogR/dlogL as calculated from the same data set,smeared with a running average over 10 points.}
\end{figure}

The daily averages of the solar irradiance have excursions between a minimum of 1362 $W/m^2$ and a maximum of 1368  $W/m^2$. The irradiance is affected by the occurrence of individual sunspots on the disk. Taking into account a 100 days running average as in fig. 2, the peak to peak amplitudes of luminosity are found to be about $\Delta L/L \sim 0.001$. The data of fig. 2 are taken from ACRIM II database\footnote{http://www.acrim.com/Data\%20Products.htm}.

\begin{figure}[top]
\centerline{\psfig{file=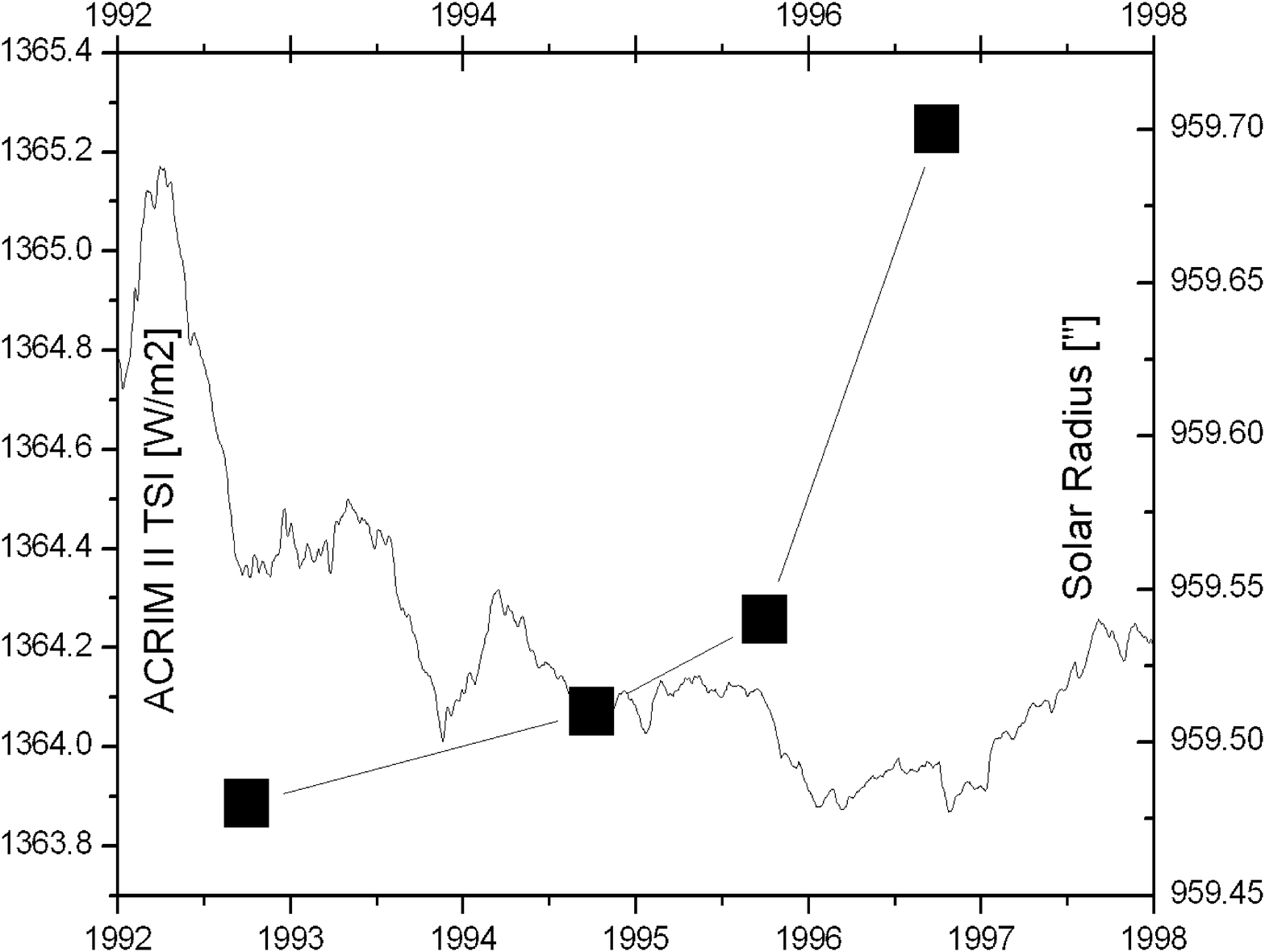,width=15cm}}
\vspace*{0pt}
\caption{The total solar irradiance (TSI) measured with ACRIM II experiment (the Active Cavity Radiometer Irradiance Monitor), available from 1992 to 1998 covering the period of SDS published data (right-hand scale and square dots).}
\end{figure}

Moreover the variations of surface temperature are limited at $1.5\pm0.2$ K and they are in phase with other indicators of the cycle\cite{Gray1997}.
Using these constraints in the equation of Stefan-Boltzmann of a radiating sphere we can set an upper limit for the variations of the solar radius.

$\Delta L/L = 2\Delta R/R + 4 \Delta T/T$

So with $\Delta L/L=0.001$ and $4 \Delta T/T =0.001$, where T=5777 K for the photosphere, and L=1365 $W/m^2$, and assuming both L and T in phase, the amplitude of  $2\Delta R/R << 0.0001$  i.e. $\Delta R << \pm 0.1''$; while assuming them in anti-phase $\Delta R \sim 0.1''$, and the maximum value allowed for radius oscillation is $\Delta R \pm 1''$ where $R=959.63''$ at 1 AU of average distance. 

It is introduced the parameter W=dlogR/dlogL in order to describe the connection between luminosity and radius. For an isothermal Sun W=0.5.
From the model plotted in fig. 1 W ranges from 0.2 to 0.6 during the main sequence phase of a solar-type star, but it is from the Picard satellite mission that this parameter is expected to be determined observationally with an unprecedented accuracy.

\section{Conclusions}

The better estimate of W is obtained with the diameters measured with SDS and the irradiance from ACRIM II satellite, as shown in fig. 2: $W=-0.3\pm0.1$. This W value is affected by the variations of TSI due to sunspots appearances on the solar disk.
The mechanism which drives TSI and solar radius changes at the 11-years timescale is different from that one which rules the evolution of the Sun at nuclear timescales represented in fig.1 (the standard model), since the two W values appear to be different.
Our measured W parameter is characteristic of the solar diameter fluctuations during the present phase of its life (under the 11 or better 22-years cycle regime\cite{Vasilyev1996} which would follow the Hale´s magnetic cycle).
When the measurements of the diameter are available, through W we get the solar luminosity in the past centuries. They will be used to feed opportune climate models.
Data on the past solar diameter (1567, 1715, 1869 and 1925) are obtained by total eclipses timings.
The Picard mission will measure this W parameter better than ever. 

\noindent {\bf Acknowledgments}
A grateful memory to the late Jean Arnaud for his continuous support to my work. Many thanks to Marianne Faurobert (Nice University) Serge Koutchmy and Cyril Bazin (Institute of Astrophysique de Paris), Sabatino Sofia, Federico Spada and Linghuai Lee (Yale University), Alexandre H. Andrei, Sergio Boscardin, Jucira Penna, and Eugenio Reis Neto (Observatorio Nacional do Rio de Janeiro), Remo Ruffini (ICRA), Michele Bianda and Renzo Ramelli (IRSOL).

\end{document}